**RESEARCH ARTICLE**  Open Access

# Getting nowhere fast: trade-off between speed and precision in training to execute image-guided hand-tool movements

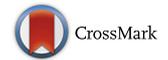

Anil Ufuk Batmaz, Michel de Mathelin and Birgitta Dresp-Langley[*]

## Abstract

**Background:** The speed and precision with which objects are moved by hand or hand-tool interaction under image guidance depend on a specific type of visual and spatial sensorimotor learning. Novices have to learn to optimally control what their hands are doing in a real-world environment while looking at an image representation of the scene on a video monitor. Previous research has shown slower task execution times and lower performance scores under image-guidance compared with situations of direct action viewing. The cognitive processes for overcoming this drawback by training are not yet understood.

**Methods:** We investigated the effects of training on the time and precision of direct view versus image guided object positioning on targets of a Real-world Action Field (RAF). Two men and two women had to learn to perform the task as swiftly and as precisely as possible with their dominant hand, using a tool or not and wearing a glove or not. Individuals were trained in sessions of mixed trial blocks with no feed-back.

**Results:** As predicted, image-guidance produced significantly slower times and lesser precision in all trainees and sessions compared with direct viewing. With training, all trainees get faster in all conditions, but only one of them gets reliably more precise in the image-guided conditions. Speed-accuracy trade-offs in the individual performance data show that the highest precision scores and steepest learning curve, for time and precision, were produced by the slowest starter. Fast starters produced consistently poorer precision scores in all sessions. The fastest starter showed no sign of stable precision learning, even after extended training.

**Conclusions:** Performance evolution towards optimal precision is compromised when novices start by going as fast as they can. The findings have direct implications for individual skill monitoring in training programmes for image-guided technology applications with human operators.

## Background

Emerging computer-controlled technologies in the biomedical and healthcare domains have created new needs for research on intuitive interactions and design control in the light of human behaviour strategies. Collecting users' views on system requirements may be a first step towards understanding how a given design or procedure needs to be adapted to better fit user needs, but is insufficient as even experts may not have complete insight into all aspects of task-specific constraints [51]. Cross-disciplinary studies focussed on interface design in the light of display ergonomics and, in priority, human psychophysics are needed to fully understand specific task environments and work domain constraints. Being able to decide what should be improved in the development and application of emerging technologies requires being able to assess how changes in design or display may facilitate human information processing during task execution. Human error [3] is a critical issue here as it is partly controlled by display properties, which may be more or less optimal under circumstances given [16, 53].

* Correspondence: birgitta.dresp@unistra.fr
Laboratoire ICube UMR 7357 CNRS-University of Strasbourg, 2, rue Boussingault, 67000 Strasbourg, France

Although there is general agreement that human cognitive processes from an integrative component of computer-assisted interventional technologies, we still do not know enough about how human performance and decision making is affected by these technologies [34]. The pressing need for research in this domain reaches far beyond the realms of workflow analysis and task models (e.g. [26]), as will be made clear here with the example of this experimental study, which addresses the problem of individual performance variations in novices learning to execute image-guided hand movements in a computer controlled simulator environment.

Image-guided interventional procedures constrain the human operator to process critical information about what his/her hands are doing in a 3D real-world environment by looking at a 2D screen representation of that environment [9]. In addition to this problem, the operator or surgeon often has to cope with uncorrected 2D views from a single camera with a fisheye lens [28, 30], providing a hemispherical focus of vision with poor off-axis resolution and aberrant shape contrast effects at the edges of the objects viewed on the screen. Novices have to learn to adapt to whatever viewing conditions, postural demands or task sequences may be imposed on them in a simulator training environment. Loss of three-dimensional vision has been pointed out as the major drawback of image-guided procedures (see [7], for a review). Compared with direct ("natural") action field viewing, 2D image viewing slows down tool-mediated task execution significantly, and also significantly affects the precision with which the task is carried out (e.g. [2, 16]). The operator or surgeon's postural comfort during task execution partly depends on where the monitor displaying the video images is placed, and there is a general consensus that it should be positioned as much as possible in line with the forearm-instrument motor axis to avoid fatigue due to axial rotation of the upper body during task execution (e.g. [7]). An off-motor-axis viewing angle of up to 45° seems to be the currently adopted standard [35]. Previously reported effects of monitor position on fatigue levels or speed of task execution [10, 20, 21, 53] point towards complex interactions between viewing angle, height of the image in the field of observation, expertise or training, and task sequencing. Varying the task sequences and allow operators to change posture between tasks, for example, was found to have significantly beneficial effects on fatigue levels of novices in simulator training for pick-and-place tasks [34].

In tool-mediated eye-hand coordination, the sensation of touch [15] is altered due to lack of haptic feed-back from the object that is being manipulated. Repeated tool-use engenders dynamic changes in cognitive hand and body schema representations (e.g. [11, 36, 37]), reflecting the processes through which highly trained experts are ultimately able to adapt to both visual and tactile constraints of image-guided interventions. Experts perform tool-mediated image-guided tasks significantly more quickly than trainees, with significantly fewer tool movements, shorter tool paths, and fewer grasp attempts [55]. Also, an expert tends to focus attention mainly on target locations, while novices split their attention between trying to focus on the targets and, at the same time, trying to track the surgical tools. This reflects a common strategy for controlling goal-directed hand movements in non-trained operators (e.g. [43]) and may affect task execution times.

Image-guided hand movements, whether mediated by a tool or not, require sensorimotor learning, an adaptive process that leads to improvement in performance through practice. This adaptive process consists of multiple distinct learning processes [29]. Hitting a target, or even getting closer to it, may generate a form of implicit reward where the trainee increasingly feels in control and where successful error reduction, which is associated with specific commands relative to the specific motor task [24], occurs naturally without external feedback. In this process, information from multiple senses (vision, touch, audition, proprioception) is integrated by the brain to generate adjustments in body, arm, or hand movements leading to faster performance with greater precision. Subjects are able to make use of error signals relative to the discrepancy between the desired and the actual movement, and the discrepancy between visual and proprioceptive estimates of body, arm, or hand positions [23, 49]. Under conditions of image-guided movement execution, real-world (direct) visual feedback is not provided, and with the unfamiliar changes in critical sensory feed-back this engenders, specific sensory integration processes may no longer be effective (see the study by [48], on the cost of expecting events in the wrong sensory modality, for example).

Here, in the light of what is summarized above, we address the problem of conditional accuracy functions in individual performance learning [38]. Conditional accuracy trade-offs occur spontaneously when novices train to perform a motor task as swiftly and as precisely as possible in a limited number of sessions [12], as is the case in laparoscopic simulator training. Conditional accuracy functions relate the duration of trial or task execution to a precision index reflecting the accuracy of the performance under conditions given [33, 41]. This relationship between speed and precision reflects hidden functional aspects of learning, and delivers important information about individual strategies the learner, especially if he/she is a beginner, is not necessarily aware of [39]. For the tutor or skill evaluator, performance trade-offs allow assessing whether a trainee is getting better at the task at hand, or whether he/she is simply getting faster without getting more precise, for example. The tutor's awareness of this kind of individual strategy problem permits intervention if necessary in the earliest phases of learning, and is essential for effective skill monitoring and for making sure that the trainee will progress in the right direction.

Surgical simulator training for image-guided interventions is currently facing the problem of defining reliable performance standards [45]. This problem partly relates to the fact that task execution time is often used as the major, or the sole criterion for establishing individual learning curves. Faster times are readily interpreted in terms of higher levels of proficiency (e.g. [54]), especially in extensive simulator training programmes hosting a large number of novice trainees. Novices are often moved from task to task in rapid succession and train by themselves in different tasks on different workstations. Times are counted by computers which generate the learning curves while the relative precision of the skills the novices are training for is, if at all, only qualitatively assessed, generally by a senior expert surgeon who himself moves from workstation to workstation. The quantitative assessment of precision requires pixel-by-pixel analyses of video image data showing hand-tool and tool-object interactions during task execution; sometimes the mechanical testing of swiftly tied knots may be necessary to assess whether they are properly tied, or come apart easily. Such analyses are costly to implement, yet, they are critically important for reasons that should become clear in the light of the findings produced in this study.

We investigated the evolution of the speed and the precision of tool-mediated (or not) and image-guided (or not) object manipulation in an object positioning task (sometimes referred to as "pick-and-place task", as for example in [34]). The task was performed by complete novices during a limited number of training sessions. In the light of previously reported data (e.g. [16]), we expect longer task execution times and lesser precision under conditions of 2D video image viewing when compared with direct ("natural") viewing. Since the experiments were run with novices, we expect tool-mediated object manipulation to be slower and less precise (e.g. [55]) when compared with bare-handed object manipulation. Previous research had shown that wearing a glove does not significantly influence task performance (e.g. [6]), but viewing conditions and tool-use were to our knowledge not included in these analyses. Here, we wanted to test whether or not wearing a glove may add additional difficulty to the already complex conditions of indirect viewing and tool-use. More importantly, we expect to observe trade-offs between task execution times and precision that are specific for each individual and can be expected to occur spontaneously (e.g. [12]) in all the training conditions, which are run without external feed-back on performance scores. The individual data of the trainees will be analyzed to bring these trade-offs to the fore and to generate conclusions relative to individual performance strategies. The implications for skill evaluation and supervised versus unsupervised simulator training will be made clear.

## Methods

Four untrained observers learned to perform the requested manual operations on an experimental simulator platform specifically designed for this purpose. This computer controlled perception-action platform (*EX-CALIBUR*) permits tracking individual task execution times in milliseconds, and an image-based analysis of task accuracy, in number of pixels, as described here below.

### Participants

Two healthy right-handed men, 25 and 27 years old, and two healthy right-handed women, 25 and 55 years old, participated in this study. Handedness was confirmed using the Edinburgh inventory for handedness designed by Oldfield [40]. The subjects were all volunteers with normal or corrected-to normal vision and naive to the purpose of the experiments. None had any experience in image-guided activities such as laparoscopic surgery training or other. Three of them stated that they did "not play videogames", one of them (subject 4) stated to "play videogames every now and again".

### Research ethics

The study was conducted in conformity with the Helsinki Declaration relative to scientific experiments on human individuals with the full approval of the ethics board of the corresponding author's host institution (CNRS). All participants were volunteers and provided written informed consent. Their identity is not revealed.

### Experimental platform

The experimental platform is a combination of hardware and software components designed to test the effectiveness of varying visual environments for image-guided action in the real world (Fig. 1). The main body of the device contains adjustable horizontal and vertical aluminium bars connected to a stable but adjustable wheel-driven sub-platform. The main body can be resized along two different axes in height and in width, and has a USB camera (ELP, Fisheye Lens, 1080p, Wide Angle) fitted into the structure for monitoring the real-world action field from a stable vertical height, which was 60 cm here in this experiment. In this study here, a single camera view was generated through one of the two 120° fisheye lens cameras, both fully adjustable in 360°, connected to a small piece of PVC. The video

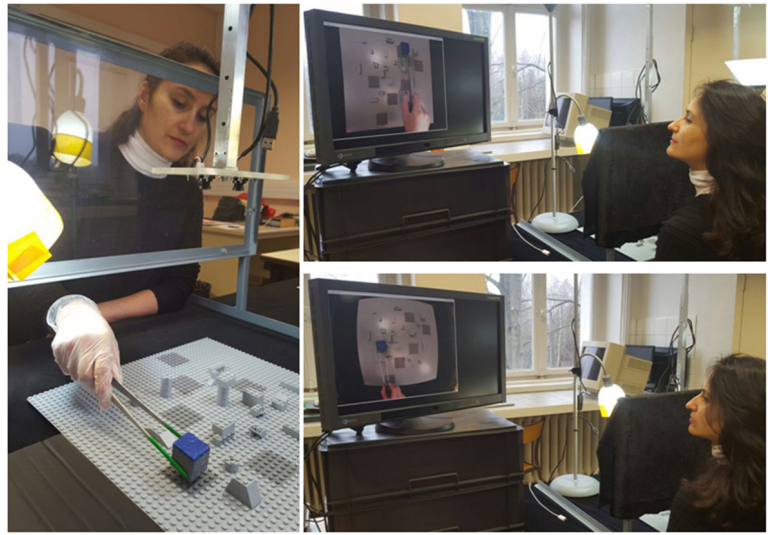

**Fig. 1** Snapshot views of the experimental platform showing experimental conditions of direct RAF viewing (*left*), 2D corrected screen viewing (*top right*), and 2D fisheye viewing (bottom right)

input received from the camera was processed by a DELL Precision T5810 model computer equipped with an Intel Xeon CPU E5-1620 with 16 Giga bytes memory (RAM) capacity at 16 bits and an NVidia GForce GTX980 graphics card. This computer is also equipped with three USB 3.0 ports, two USB 2.0 SS ports and two HDMI video output generators. The operating system uses Windows 7. Experiments are programmed in Python 2.7 using the Open CV computer vision software library. The computer was connected to a high resolution color monitor (EIZO LCD 'Color Edge CG275W') with an in-built color calibration device (colorimeter), which uses the Color Navigator 5.4.5 interface for Windows. The colors of objects visualized on the screen can be matched to LAB or RGB color space, fully compatible with Photoshop 11 and similar software tools. The color coordinates for RGB triples can be retrieved from a look-up table at any moment in time after running the auto-calibration software.

### Objects in the real-world action field

The Real-world Action Field (as of now referred to as the RAF) consisted of a classic square shaped (45 cm × 45 cm) light grey LEGO© board available worldwide in the toy sections of large department stores. Six square-shaped (4,5 cm × 4,5 cm) target areas were painted on the board at various locations in a medium grey tint (acrylic). In-between these target areas, small LEGO© pieces of varying shapes and heights were placed to add a certain level of complexity to both the visual configuration and the task and to reduce the likelihood of getting performance ceiling effects. The object that had to be placed on the target areas in a specific order was a small (3 cm × 3 cm × 3 cm) cube made of very light plastic foam but resistant to deformation in all directions. Five sides of the cube were painted in the same medium grey tint (acrylic) as the target areas. One side, which was always pointing upwards in the task (Fig. 1, image on left), was given an ultramarine blue tint (acrylic) to permit tracking object positions. A medium sized barbecue tong with straight ends was used for manipulating the object in the conditions 'with tool' (Fig. 1, image on left). The tool-tips were given a matte fluorescent green tint (acrylic) to permit tool-tip tracking. The surgical gloves used in the conditions 'with glove' (Fig. 1, image on left) were standard, medium size surgical vinyl gloves available in pharmacies.

### Objects visualized on screen

The video input received by the computer from the USB camera generates raw image data within a viewing frame of the dimensions 640 pixels (width) × 480 pixels (height). These data were processed to generate show image data in a viewing frame of the dimensions 1280 pixels (width) × 960 pixels (height), the size of a single pixel on the screen being 0.32 mm. The size of the RAF (grey LEGO© board) visualized on the computer screen was identical to that in the real world (45 cm × 45 cm), and so were the size of the target areas (4,5 cm × 4,5 cm) and of the object manipulated (3 cm × 3 cm). A camera output matrix with image distortion coefficients using the Open CV image library in Python was used to correct the fisheye effects for the 2D corrected viewing conditions of the experiment. This did not affect the size dimensions of the visual objects given here above. The luminance ($L$) of the light grey RAF visualized on the screen was 33,8 cd/m$^2$ and the luminance of the medium grey target areas was 15,4 cd/m², producing a target/background contrast (Weber contrast: $((L_{foreground}-L_{background})/L_{background}))$ of -0,54. The luminance of the blue (x = 0,15, y = 0,05, z = 0,80 in CIE color space) object surface visualized on the screen was 3,44 cd/m², producing Weber contrasts of −0,90 with regard to the RAF, and −0,78 with regard to the target areas. The luminance (29,9 cd/m²) of the green (x = 0,20, y = 0,70, z = 0,10 in CIE color space) tool-tips produced Weber contrasts of −0,11 with regard to the RAF, and 0,94 with regard to the target areas. All luminance values for calculating the object contrasts visualized on the screen were obtained on the basis of standard photometry using an external photometer (Cambridge Research Instruments) with the adequate interface software. These calibrations were necessary to ensure that the image conditions matched the direct viewing condition as closely as possible. Temporal matching was controlled by the algorithm driving the internal clock of the CPU, ensuring that the video-images where synchronized with the real-world actions.

## Experimental design

A Cartesian design plan $P_{4x}T_{2}xV_{3}xM_{2}xS_{8}$ was adopted for testing the expected effects of training, viewing modality, and object manipulation mode on inter-individual variations in time and precision during training, specified here above in the last paragraph of the introduction. To this purpose, four participants ($P_4$) performed the experimental task in three ('direct' vs 'fisheye' vs 'corrected 2D') viewing conditions ($V_3$) and two conditions ('with tool' vs 'without tool') of object manipulation ($M_2$), and two modalities ('bare hand' vs 'glove') of touch ($T_2$) in eight successive training sessions ($S_8$). The order of conditions was counterbalanced between participants and sessions (see experimental procedure here below). There were ten repeated trial sets for each combination of conditions within a session, yielding a total of 3840 experimental observations for 'time' and for 'precision'.

## Procedure

The experiments were run under conditions of free viewing, with general illumination levels that can be assimilated to daylight conditions. The RAF was illuminated by two lamps (40Watt, 6500 K), constantly lit during the whole duration of the experiment. Participants were comfortably seated at a distance of approximately 75 cm from the RAF in front of them, and from the screen, which was positioned at an angle of slightly less than 45° to their left. As explained in the introduction, this monitor position is within the range of currently accepted standards for comfort. A printout of the targets-on-RAF configuration was handed out to the participant at the beginning. White straight lines on the printout indicated the ideal object trajectory, and red numbers indicated the order in which the small blue cube object had to be placed on the light grey targets in a given trial set (Fig. 2). The pick-and-place sequence was always from position zero to position one, then to two, to three, to four, to five, then back to position zero. Participants were instructed to position the cube with their dominant hand "as precisely as possible and as swiftly as possible on the center of each target, in the right order as indicated on the printout". They were also informed that they were going to perform this task under different conditions of object manipulation: with and without a tool, with their bare hands and wearing a surgical glove, while viewing the RAF (and their own hands) in front of them, and while viewing the RAF (and their own hands) on a computer screen. In the direct viewing condition, participants saw the RAF and what their hands were doing through a glass window, which was covered by a black velvet curtain. In the 2D video conditions, subjects saw an image of the RAF on the computer screen. All participants grasped the object with the thumb and the index of their right hand, from the same angle, when no tool was used. When using the tool, they all had to approach the object from the front to grasp it with the two tool-tips. Before starting the first trial set, the participant could look at the printout of the task trajectory for as long as he/she wanted. When they felt confident that they remembered the target order well enough to do the task, the printout was taken away from them. An individual experiment was always started with a "warm-up" run in each of the different conditions. Data were collected from the moment a participant was able to produce a trial sequence without missing the target area or dropping the object. An experimental session always began with the easiest

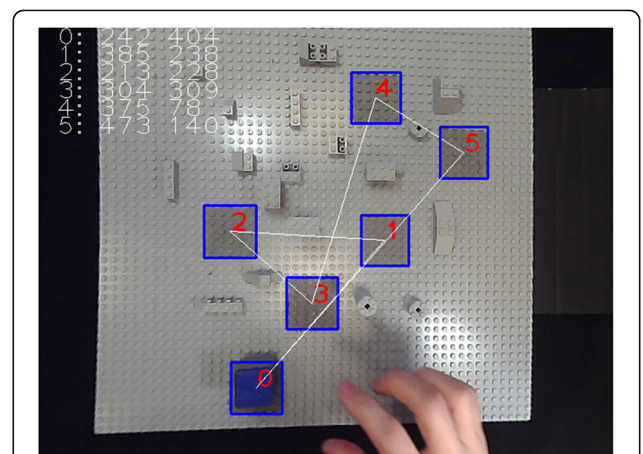

**Fig. 2** Screenshot view of the RAF, with the ideal object trajectory, from position zero to the positions one, two, three, four, five, and back to zero. Participants had to position a small foam cube with a blue top on the centers of the grey target areas in the right order as precisely as possible and as swiftly as possible

(cf. [16]) condition of direct viewing. Thereafter the order of the two 2D viewing conditions (2D corrected and 2D fisheye) was counterbalanced, between sessions and between participants, to avoid order specific habituation effects. For the same reason, the order of the tool-use conditions (with and without tool) and the touch conditions (with and without glove) was also counterbalanced, between sessions and between participants. No performance feed-back was given. At the end of training, each participant was able to see his/her learning curves from the eight sessions, for both 'time' and 'precision'. No specific comments were communicated to them, and no questions were asked at this stage. Subject 4 spontaneously wanted to run in twelve additional sessions to see whether he could produce any further evolution in his performance.

### Data generation

Data from fully completed trial sets only were recorded. A fully complete trial set consists of a set of positioning operations starting from zero, then going to one, to two, to three, to four, to five, and back to position zero without dropping the object accidentally and without errors in the positioning order. Whenever such occurred (this happened only incidentally, mostly at the beginning of the experiment), the trial set was aborted immediately and the participant started from scratch in that specific condition.

Ten fully completed trial sets were recorded for each combination of factor levels. For each of such ten trial sets, the computer program generated data relative to the dependent variables 'time' and 'precision'. For 'time', the computer program counts the CPU time (in milliseconds) from the moment the blue cube object is picked up by the participant to the time it is put back to position zero again. The rate for image-time data collection is between 25 and 30 Hz, with an error margin of less than 40 milliseconds for any of the time estimates. For 'precision', the computer program counts the number of blue object pixels at positions "off" the 3 cm × 3 cm central area of each of the five 4,5 cm × 4,5 cm target areas (see Fig. 3) whenever the object is positioned on a target. The standard error of these positional estimates, determined in the video-image calibration procedure, was always smaller than 10 pixels. "Off"-center pixels were not counted for object positions on the square labeled 'zero' (the departure and arrival square). Individual time and precision data were written to an excel file by the computer program, with labeled data columns for the different conditions, and stored in a directory for subsequent analysis.

### Results

The data recorded from each of the subjects were analyzed as a function of the different experimental conditions, for each of the two dependent variables ('time' and 'precision'). Medians and scatter of the individual distributions relative to 'time' and 'precision' were computed first. Box-and-whiskers plots were generated to visualize these distributions. Means and their standard errors for 'time' and 'precision' were computed in the next step, for each subject and experimental condition. The raw data were submitted to analysis of variance (ANOVA) and conditional plots of means and standard errors as a function of the rank number of the trial sessions were generated for each subject to show the evolution of 'time' and 'precision' with training.

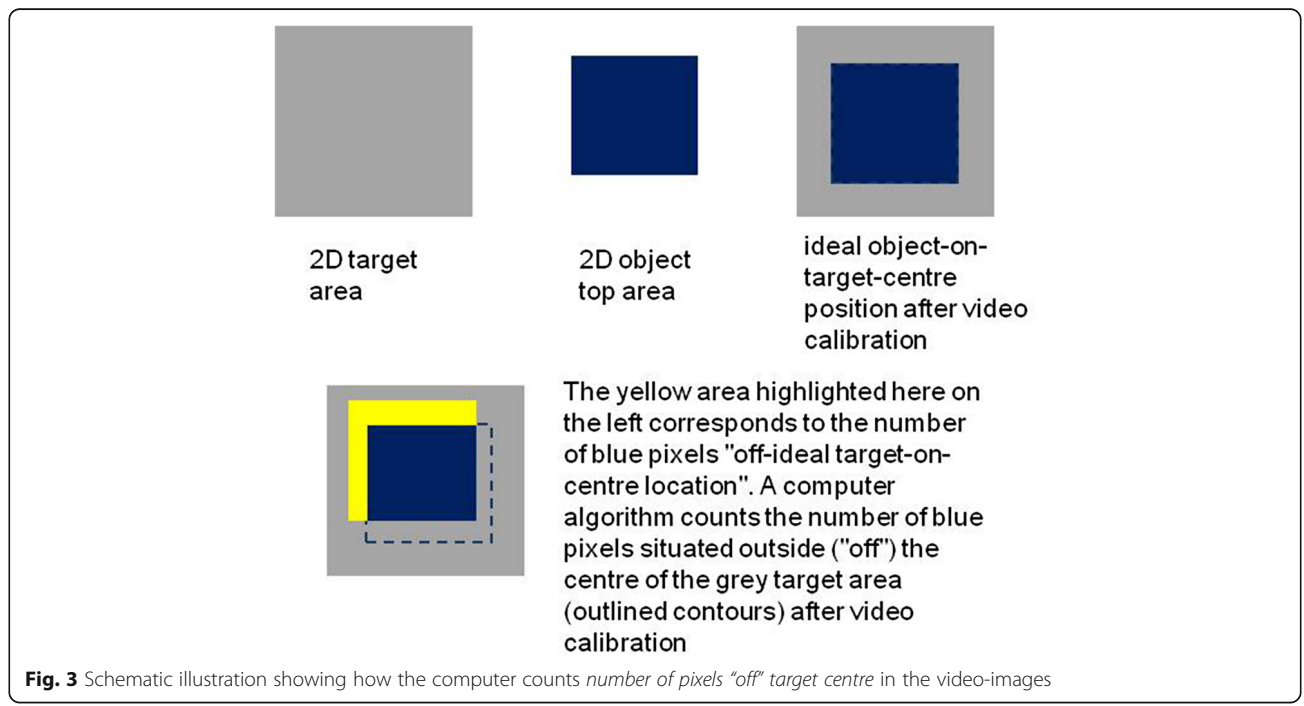

**Fig. 3** Schematic illustration showing how the computer counts *number of pixels "off" target centre* in the video-images

## Medians and extremes

Medians and extremes of the individual data relative 'time' and 'precision' for the different experimental conditions were analyzed first. The results of this analysis are represented graphically as box-and-whiskers plots here in Figs. 4 and 5. Figure 4 shows distributions around the medians of data from the manipulation modality with tool in the three different viewing conditions. Figure 5 shows distributions around the medians of data from the manipulation modality without tool in the three different viewing conditions. The distributions around the medians, with upper and lower extremes, for the data relative to 'time' show that Subject 1 was the slowest in all conditions, closely followed by Subject 2. Subjects 3 and 4 were noticeably faster in all conditions and their distributions for 'time' generally display the least scatter around the median. All subjects took longer in the tool-mediated manipulation modality (see graphs on left in Fig. 4) compared with the by-hand manipulation modality without tool. The shortest times are displayed in the distributions from the direct viewing condition and the longest times in the distributions from the fisheye image viewing condition. Medians, upper and lower quartiles and extremes for 'precision' (graphs on right) show that subject 1 is the most precise in all conditions, with distributions displaying the smallest number of pixels "off" target center and the least scatter around the medians. Subject 2 was the least precise, with distributions displaying the largest number of pixels "off" target center and the most scatter around the medians in most conditions except in the direct viewing conditions without tool, where subject 3´s distribution displays the largest "off" center values and the least scatter around the median. All other subjects were the most precise in the direct viewing conditions, excluding the two outlier data points at the upper extremes of the distributions of subject 3 and 4. Subject 2 was the least precise in the fisheye image viewing conditions, and the

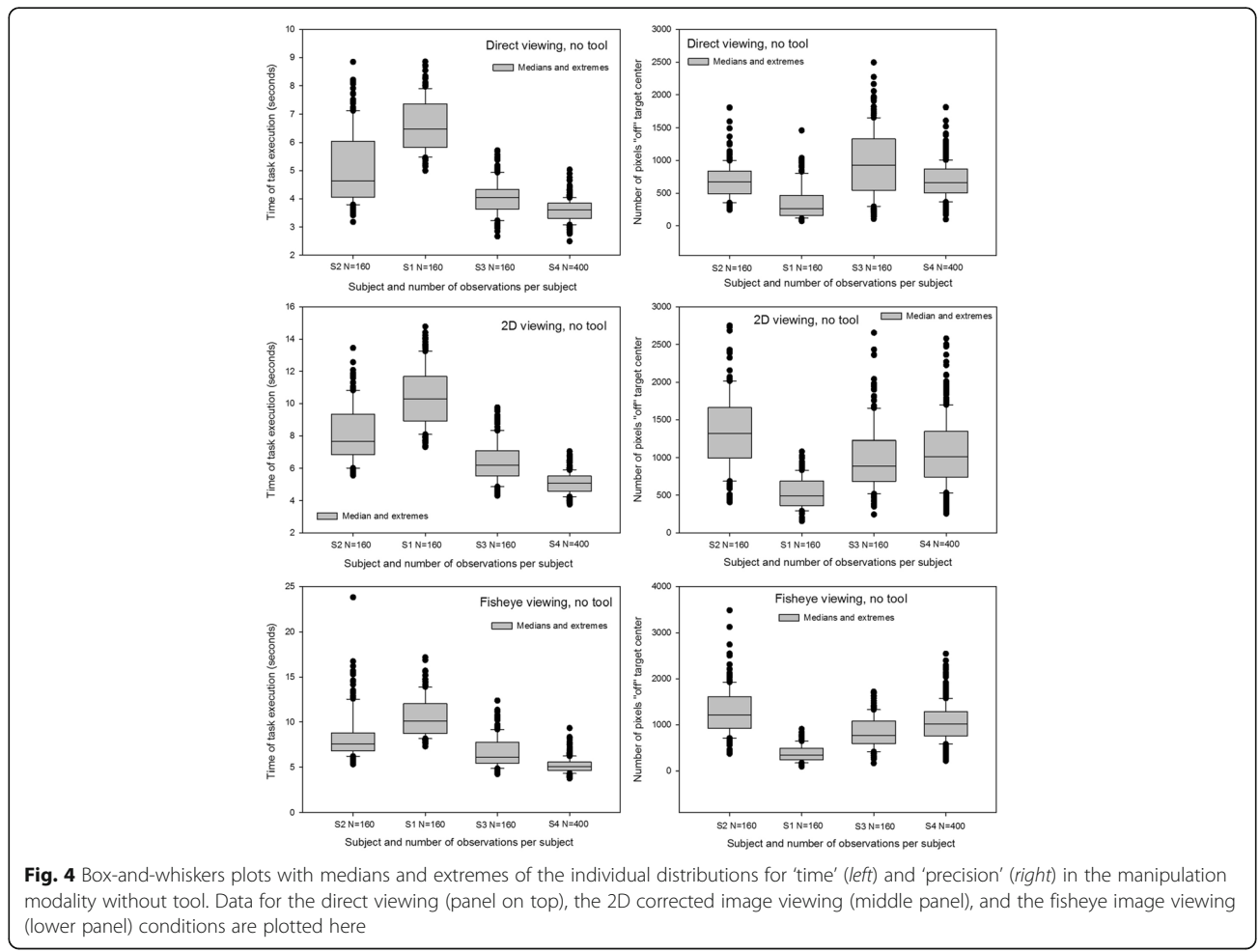

**Fig. 4** Box-and-whiskers plots with medians and extremes of the individual distributions for 'time' (*left*) and 'precision' (*right*) in the manipulation modality without tool. Data for the direct viewing (panel on top), the 2D corrected image viewing (middle panel), and the fisheye image viewing (lower panel) conditions are plotted here

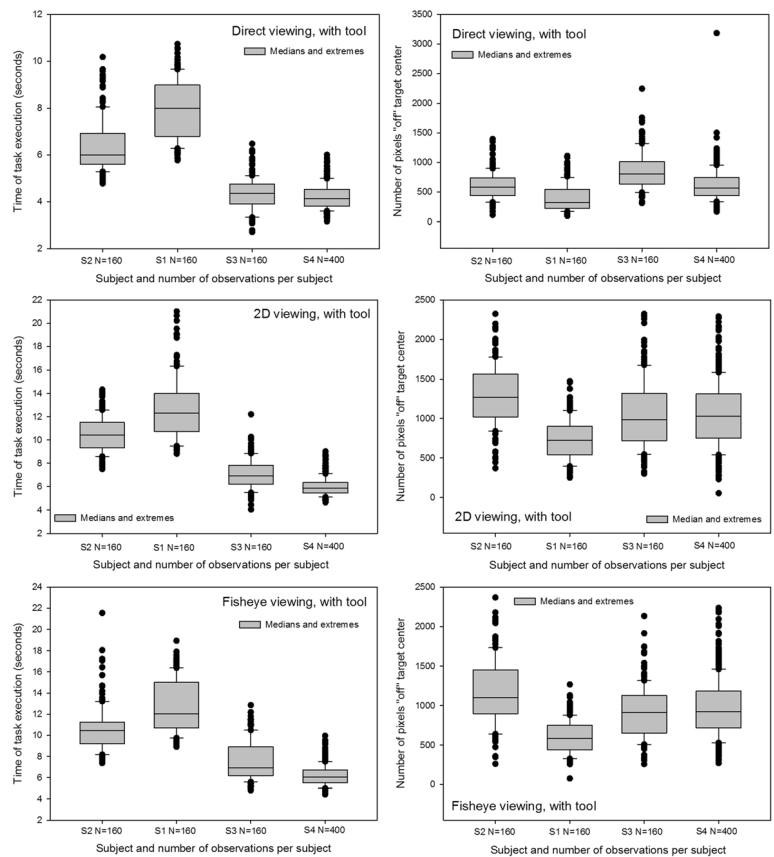

**Fig. 5** Box-and-whiskers plots with medians and extremes of the individual distributions for 'time' (*left*) and 'precision' (*right*) in the manipulation modality with tool, for the direct viewing (*upper panel*), the 2D corrected image viewing (*middle panel*), and the fisheye image viewing (*lower panel*) conditions

three other subjects were the least precise in the 2D corrected image viewing conditions.

## Analysis of variance
Two outliers at the upper extremes of the distributions around the medians relative to 'time' of subject 2 in the fisheye viewing conditions with and without tool, and two outliers at the upper extremes of the distributions around the medians relative to 'precision' of subjects 4 and 5 in the direct viewing condition without tool were corrected by replacing them by the mean of the distribution. 3840 raw data for 'time' and 3840 raw data for 'precision' were submitted to Analysis of Variance (ANOVA) in *MATLAB 7.14*. The distributions for 'time' and 'precision' satisfy general criteria for parametric testing (independence of observations, normality of distributions and equality of variance). 5-Way ANOVA was performed for a design plan $P_4xT_2xV_3xM_2xS_8$ with four levels of the 'participant' factor $P_4$, which is analyzed as a main experimental factor here because we are interested in differences between individuals, as explained earlier in the introduction and the *experimental design* paragraph.

## Principal variables
The differences between means for 'time' and 'precision' of the different levels of each factor were statistically significant for almost all experimental factors except for effects of 'touch' $T_2$ on 'time' and effects of 'manipulation' $M_2$ on 'precision'. Means (M) and standard errors (SEM) for each level of each principal variable, and the ANOVA results, with $F$ values and the associated degrees of freedom and probability limits, are summarized in Table 1. The differences between means for 'time' and 'precision' of the three levels of the 'viewing' factor displayed in the table show that participants were significantly slower and significantly less precise in the image guided conditions compared with the direct viewing condition. Comparing the means for the two levels of 'manipulation' ($M_2$) shows that tasks were executed significantly faster when no tool was used, with no significant difference in precision. The 'touch' factor($T_2$) had no effect on task execution times, but participants were significantly less precise when wearing a glove. The most critical factors for our learning study here, the 'session' ($S_8$) and 'participant' ($P_4$) factors, produced

**Table 1** 5-Way *ANOVA* summary

| Factor | Factor Level | M | SEM | ANOVA on time | M | SEM | ANOVA on precision |
|---|---|---|---|---|---|---|---|
| Viewing | Direct | 5.34 | 0.05 | $F_{(2,3839)} = 9953.73 p < .001$ | 636 | 9 | $F_{(2,3839)} = 509.26 p < .001$ |
| | Fisheye | 8.65 | 0.08 | | 886 | 12 | |
| | 2D | 8.42 | 0.08 | | 995 | 13 | |
| Manipulation | Tool | 8.16 | 0.08 | $F_{(1,3839)} = 4176.32 p < .001$ | 841 | 9 | $F_{(1,3839)} = 0.28;$ NS |
| | No Tool | 6.78 | 0.06 | | 836 | 11 | |
| Touch | Glove | 7.47 | 0.07 | $F_{(1, 3839)} = 0.05$ NS | 850 | 20 | $F_{(2,3839)} = 5.86; p < .02$ |
| | No Glove | 7.47 | 0.07 | | 827 | 10 | |
| Session | Session 1 | 9.52 | 0.18 | $F_{(7,3839)} = 1452.22 p < .001$ | 900 | 23 | $F_{(2,3839)} = 56.79; p < .001$ |
| | … | … | … | | … | … | |
| | Session 8 | 6.62 | 0.1 | | 745 | 19 | |
| Participant | 1 | 10.17 | 0.07 | $F_{(3,3839)} = 9986.77 p < .001$ | 503 | 9 | $F_{(2,3839)} = 653.91 p < .001$ |
| | 2 | 8.17 | 0.05 | | 1065 | 14 | |
| | 3 | 6.03 | 0.04 | | 938 | 11 | |
| | 4 | 5.53 | 0.03 | | 950 | 11 | |

Summary of main results of the 5-Way ANOVA. Means (M) for the dependent variables 'time' (left) and 'precision' (right) and their standard errors (SEM) are given for the different levels of each principal variable (factor). The F values, with degrees of freedom and probabilities limits, for the effect of each factor on each dependent variable are shown significant effects on 'time' and on 'precision'. These can, however, not be summarized without taking into account their interaction, which was significant for 'time' (F (21, 3839) = 162.88; $p < .001$) and for 'precision' (F (21, 3839) = 35.21; $p < .001$).

### Interactions

The 'participant' and 'session' factors produced significant interactions with the 'viewing' factor: (F(14, 3839) = 104.67; $p < .001$ for 'session' x 'viewing' on 'time' and F(6, 3839) = 267.74; $p < .001$ for 'participant' x 'viewing' on 'time'; (F(14, 3839) = 3.86; $p < .001$ for 'session' x 'viewing' on 'precision' and F(6, 3839) = 81.32; $p < .001$ for 'participant' x 'viewing' on 'precision'. To further quantify these complex interactions, *post-hoc* comparisons (Holm-Sidak procedure, the most robust for this purpose) for the three levels of 'viewing' ($V_3$) and the eight levels of 'session' ($S_8$) in each level (p1, p2, p3, and p4) of the 'participant' factor ($P_4$) were carried out for both dependent variables. The degrees of freedom (*df*) of these step-down tests are *N-k*, where *N* is the sample size (here 3840/4 = 960) and *k* the number of factor levels (here 3 + 8 = 12) compared in each test. The results of these *post-hoc* comparisons are displayed in Tables 2, 3, 4, 5, 6, 7, 8 and 9, which give effect sizes in terms of differences in means, for 'time' and 'precision', between the viewing conditions for each participant and session, *t* values, and the corresponding unadjusted probabilities. In these tables we see that the effect sizes do not evolve in the same way in the different participants as the sessions progress.

In the next step of the analysis, the conditional data for 'time' and 'precision' were represented graphically.

Figure 6 shows the effects of 'session' ($S_8$) on 'time' (left) and on 'precision' (right). Figure 7 shows the effects of 'participant' ($P_4$) on 'time' (left) and 'precision' (right). For further insight into differences between participants, their individual functions (means and standard errors of the conditional performance scores) were plotted as a function of the rank number of the sessions. These functions permit tracking the evolution of individual performance with training

### Individual performance evolution with training

These individual data are plotted in Fig. 8 (data of subject 1, female), Fig. 9 (subject 2´s data, female), Fig. 10 (subject 3´s data, male) and Fig. 11 (subject 4´s data, male). The upper figure panels show average data for 'time' and 'precision' as a function of the rank number of the training session, the lower panels show the corresponding standard errors (SEM). Comparisons between individuals show that subject 1 starts with the slowest times, while the other three participants start noticeably faster, especially subjects 3 and 4, with subject 4 being the fastest of all. Subject 1, while being the slowest of all, starts with the best performance in precision, with the smallest "off" target pixel score, and keeps getting more precise with training while getting faster at the same time. Her precision levels in the last of her eight training sessions are the best compared with the three others, with the smallest standard errors in all the training sessions. Her times at the end of training are comparable with the times of subject 2 at the beginning of the sessions, who gets faster thereafter but, at the same time, is the least accurate and does not get any better in the

**Table 2** *Post-hoc* comparisons - effects on time in participant 1

|  | D Means | t | P |
|---|---|---|---|
| Session 1 | | | |
|   2D vs. Direct | 6.772 | 28.07 | 0.000 |
|   2D vs. Fisheye | 0.040 | 0.17 | 0.867 NS |
|   Fisheye vs. Direct | 6.732 | 27.91 | 0.000 |
| Session 2 | | | |
|   2D vs. Direct | 5.231 | 21.69 | 0.000 |
|   2D vs. Fisheye | 0.440 | 1.82 | 0.068 NS |
|   Fisheye vs. Direct | 5.671 | 23.51 | 0.000 |
| Session 3 | | | |
|   2D vs. Direct | 3.752 | 15.55 | 0.000 |
|   2D vs. Fisheye | 1.145 | 4.75 | 0.000 |
|   Fisheye vs. Direct | 4.897 | 20.30 | 0.000 |
| Session 4 | | | |
|   2D vs. Direct | 3.721 | 15.43 | 0.000 |
|   2D vs. Fisheye | 0.677 | 2.81 | 0.005 |
|   Fisheye vs. Direct | 3.045 | 12.62 | 0.000 |
| Session 5 | | | |
|   2D vs. Direct | 4.381 | 18.16 | 0.000 |
|   2D vs. Fisheye | 0.492 | 2.04 | 0.041 |
|   Fisheye vs. Direct | 3.889 | 16.12 | 0.000 |
| Session 6 | | | |
|   2D vs. Direct | 4.940 | 20.48 | 0.000 |
|   2D vs. Fisheye | 0.117 | 0.48 | 0.682 NS |
|   Fisheye vs. Direct | 4.823 | 19.99 | 0.000 |
| Session 7 | | | |
|   2D vs. Direct | 2.660 | 11.03 | 0.000 |
|   2D vs. Fisheye | 0.296 | 1.23 | 0.219 NS |
|   Fisheye vs. Direct | 2.956 | 12.26 | 0.000 |
| Session 8 | | | |
|   2D vs. Direct | 3.032 | 12.57 | 0.000 |
|   2D vs. Fisheye | 0.048 | 0.19 | 0.843 NS |
|   Fisheye vs. Direct | 2.984 | 12.37 | 0.000 |

Results of the *post-hoc* comparisons for effects on time of the three levels of *'viewing'* ($V_3$) in the eight levels of *'session'* ($S_8$) in level 1 of the *'participant'* factor. Effect sizes (D Means), t values, and unadjusted probabilities (P) are given for each comparison

**Table 3** *Post-hoc* comparisons - effects on precision in participant 1

|  | D Means | t | P |
|---|---|---|---|
| Session 1 | | | |
|   2D vs. Direct | 610.9 | 8.91 | 0.000 |
|   2D vs. Fisheye | 161.5 | 2.38 | 0.020 |
|   Fisheye vs. Direct | 461.5 | 5.91 | 0.000 |
| Session 2 | | | |
|   2D vs. Direct | 300.3 | 4.34 | 0.000 |
|   2D vs. Fisheye | 147.1 | 2.13 | 0.027 |
|   Fisheye vs. Direct | 153.2 | 2.21 | 0.033 |
| Session 3 | | | |
|   2D vs. Direct | 468.8 | 6.77 | 0.000 |
|   2D vs. Fisheye | 173.9 | 2.51 | 0.012 |
|   Fisheye vs. Direct | 294.8 | 4.26 | 0.000 |
| Session 4 | | | |
|   2D vs. Direct | 8.8 | 0.17 | 0.126 NS |
|   2D vs. Fisheye | 11.9 | 0.24 | 0.102 NS |
|   Fisheye vs. Direct | 30.2 | 0.92 | 0.095 NS |
| Session 5 | | | |
|   2D vs. Direct | 366.5 | 5.30 | 0.000 |
|   2D vs. Fisheye | 218.4 | 3.15 | 0.002 |
|   Fisheye vs. Direct | 140.4 | 2.04 | 0.032 |
| Session 6 | | | |
|   2D vs. Direct | 29.8 | 0.76 | 0.222 NS |
|   2D vs. Fisheye | 56.4 | 0.43 | 0.201 NS |
|   Fisheye vs. Direct | 83.4 | 0.84 | 0.098 NS |
| Session 7 | | | |
|   2D vs. Direct | 50.3 | 1.25 | 0.133 NS |
|   2D vs. Fisheye | 41.2 | 0.19 | 0.224 NS |
|   Fisheye vs. Direct | 240.0 | 3.81 | 0.002 |
| Session 8 | | | |
|   2D vs. Direct | 80.5 | 1.06 | 0.089 NS |
|   2D vs. Fisheye | 56.5 | 0.31 | 0.156 NS |
|   Fisheye vs. Direct | 66.0 | 0.13 | 0.222 NS |

Results of the *post-hoc* comparisons for effects on precision of the three levels of *'viewing'* ($V_3$) in the eight levels of *'session'* ($S_8$) in level 1 of the *'participant'* factor. Effect sizes (D Means), t values, and unadjusted probabilities (P) are given for each comparison eight training sessions. Subjects 3 and 4 both start with the fastest times. Subject 3´s precision first improves drastically in the first session, then gets worse again as he is getting faster. In the last sessions, this subject's performance improves with regard to precision while the times and their standard errors remain stable. Subject 4 is the fastest performer. His average times and their standard errors decrease steadily with training and level off at the lowest level after his eight first training sessions. Precision, however, does not evolve, but varies considerably in all the training sessions, with the highest standard errors. Adding another 12 training sessions for this subject results in even faster performances in all conditions with even lower standard errors, however, precision does not improve noticeably in any of the image viewing conditions, it improves a little in the direct viewing condition when a tool is used to execute the object positioning task. All subjects perform best, and improve to a greater or lesser extent in time and

**Table 4** *Post-hoc* comparisons - effects on time in participant 2

|  | D Means | t | P |
|---|---|---|---|
| Session 1 |  |  |  |
|   2D vs. Direct | 3.709 | 15.37 | 0.000 |
|   2D vs. Fisheye | 2.667 | 11.06 | 0.000 |
|   Fisheye vs. Direct | 6.376 | 26.43 | 0.000 |
| Session 2 |  |  |  |
|   2D vs. Direct | 4.887 | 20.26 | 0.000 |
|   2D vs. Fisheye | 1.423 | 5.89 | 0.000 |
|   Fisheye vs. Direct | 3.464 | 14.36 | 0.000 |
| Session 3 |  |  |  |
|   2D vs. Direct | 3.249 | 13.47 | 0.000 |
|   2D vs. Fisheye | 0.330 | 1.37 | 0.171 NS |
|   Fisheye vs. Direct | 3.579 | 14.84 | 0.000 |
| Session 4 |  |  |  |
|   2D vs. Direct | 3.632 | 15.06 | 0.000 |
|   2D vs. Fisheye | 0.923 | 3.82 | 0.000 |
|   Fisheye vs. Direct | 2.710 | 11.23 | 0.000 |
| Session 5 |  |  |  |
|   2D vs. Direct | 2.639 | 10.94 | 0.000 |
|   2D vs. Fisheye | 0.706 | 2.93 | 0.000 |
|   Fisheye vs. Direct | 3.345 | 13.87 | 0.000 |
| Session 6 |  |  |  |
|   2D vs. Direct | 2.512 | 10.41 | 0.000 |
|   2D vs. Fisheye | 0.278 | 1.15 | 0.250 NS |
|   Fisheye vs. Direct | 2.234 | 9.26 | 0.000 |
| Session 7 |  |  |  |
|   2D vs. Direct | 4.112 | 17.05 | 0.000 |
|   2D vs. Fisheye | 0.249 | 1.03 | 0.302 NS |
|   Fisheye vs. Direct | 4.361 | 18.08 | 0.000 |
| Session 8 |  |  |  |
|   2D vs. Direct | 4.069 | 16.87 | 0.000 |
|   2D vs. Fisheye | 0.072 | 0.29 | 0.765 NS |
|   Fisheye vs. Direct | 3.997 | 16.57 | 0.000 |

Results of the *post-hoc* comparisons for effects on time of the three levels of 'viewing' ($V_3$) in the eight levels of 'session' ($S_8$) in level 2 of the 'participant' factor. Effect sizes (D Means), t values, and unadjusted probabilities (P) are given for each comparison

**Table 5** *Post-hoc* comparisons - effects on precision in participant 2

|  | D Means | t | P |
|---|---|---|---|
| Session 1 |  |  |  |
|   2D vs. Direct | 764.1 | 11.01 | 0.000 |
|   2D vs. Fisheye | 435.7 | 6.28 | 0.000 |
|   Fisheye vs. Direct | 328.3 | 4.73 | 0.000 |
| Session 2 |  |  |  |
|   2D vs. Direct | 787.2 | 9.26 | 0.000 |
|   2D vs. Fisheye | 524.0 | 7.55 | 0.000 |
|   Fisheye vs. Direct | 263.2 | 3.09 | 0.000 |
| Session 3 |  |  |  |
|   2D vs. Direct | 432.5 | 5.09 | 0.004 |
|   2D vs. Fisheye | 199.2 | 2.88 | 0.000 |
|   Fisheye vs. Direct | 622.8 | 8.96 | 0.000 |
| Session 4 |  |  |  |
|   2D vs. Direct | 768.7 | 11.81 | 0.000 |
|   2D vs. Fisheye | 26.8 | 0.38 | 0.698 NS |
|   Fisheye vs. Direct | 741.2 | 10.71 | 0.000 |
| Session 5 |  |  |  |
|   2D vs. Direct | 741.3 | 11.02 | 0.000 |
|   2D vs. Fisheye | 198.1 | 2.88 | 0.004 |
|   Fisheye vs. Direct | 563.2 | 8.35 | 0.010 |
| Session 6 |  |  |  |
|   2D vs. Direct | 535.2 | 6.29 | 0.000 |
|   2D vs. Fisheye | 31.1 | 0.45 | 0.653 NS |
|   Fisheye vs. Direct | 588.0 | 5.59 | 0.000 |
| Session 7 |  |  |  |
|   2D vs. Direct | 558.3 | 6.57 | 0.000 |
|   2D vs. Fisheye | 110.4 | 1.59 | 0.111 NS |
|   Fisheye vs. Direct | 442.2 | 6.39 | 0.000 |
| Session 8 |  |  |  |
|   2D vs. Direct | 890.3 | 12.10 | 0.000 |
|   2D vs. Fisheye | 262.3 | 3.07 | 0.002 |
|   Fisheye vs. Direct | 528.0 | 8.34 | 0.007 |

Results of the *post-hoc* comparisons for effects on precision of the three levels of 'viewing' ($V_3$) in the eight levels of 'session' ($S_8$) in level 2 of the 'participant' factor. Effect sizes (D Means), t values, and unadjusted probabilities (P) are given for each comparison precision of task execution in the direct viewing conditions. In the fisheye image viewing and the corrected 2D viewing conditions, only the performances of subject 1 and subject 3 become more accurate with training. Subject 2´s precision gets worse rather than better with training in the image viewing conditions. Subject 4´s precision remains unstable, with highs and lows up to the last of his twenty training sessions, where his average times and their standard errors have leveled out at the best possible performance score for 'time' under the task conditions given.

## Discussion

As would be expected on the basis of previous observations [2, 7, 16], our results confirm that 2D video-image viewing negatively affects both time and precision of task execution compared with direct action viewing (control). This performance loss is statistically significant. Although

**Table 6** *Post-hoc* comparisons - effects on time in participant 3

|  | D Means | t | P |
|---|---|---|---|
| **Session 1** |  |  |  |
| 2D vs. Direct | 3.442 | 14.27 | 0.000 |
| 2D vs. Fisheye | 1.820 | 7.55 | 0.000 |
| Fisheye vs. Direct | 5.263 | 21.82 | 0.000 |
| **Session 2** |  |  |  |
| 2D vs. Direct | 3.800 | 15.76 | 0.000 |
| 2D vs. Fisheye | 0.952 | 3.95 | 0.000 |
| Fisheye vs. Direct | 4.753 | 19.70 | 0.000 |
| **Session 3** |  |  |  |
| 2D vs. Direct | 2.998 | 12.43 | 0.000 |
| 2D vs. Fisheye | 0.423 | 1.75 | 0.079 NS |
| Fisheye vs. Direct | 3.421 | 14.18 | 0.000 |
| **Session 4** |  |  |  |
| 2D vs. Direct | 2.150 | 8.91 | 0.000 |
| 2D vs. Fisheye | 0.039 | 0.16 | 0.870 NS |
| Fisheye vs. Direct | 2.189 | 9.07 | 0.000 |
| **Session 5** |  |  |  |
| 2D vs. Direct | 1.581 | 6.55 | 0.000 |
| 2D vs. Fisheye | 0.325 | 1.35 | 0.178 NS |
| Fisheye vs. Direct | 1.906 | 7.90 | 0.000 |
| **Session 6** |  |  |  |
| 2D vs. Direct | 2.146 | 8.89 | 0.000 |
| 2D vs. Fisheye | 0.094 | 0.39 | 0.694 NS |
| Fisheye vs. Direct | 2.051 | 8.50 | 0.000 |
| **Session 7** |  |  |  |
| 2D vs. Direct | 2.360 | 9.78 | 0.000 |
| 2D vs. Fisheye | 0.257 | 1.06 | 0.288 NS |
| Fisheye vs. Direct | 2.103 | 8.72 | 0.000 |
| **Session 8** |  |  |  |
| 2D vs. Direct | 1.958 | 8.17 | 0.000 |
| 2D vs. Fisheye | 0.124 | 0.51 | 0.608 NS |
| Fisheye vs. Direct | 1.834 | 7.61 | 0.000 |

Results of the *post-hoc* comparisons for effects on time of the three levels of 'viewing' ($V_3$) in the eight levels of 'session' ($S_8$) in level 3 of the 'participant' factor. Effect sizes (D Means), t values, and unadjusted probabilities (P) are given for each comparison

**Table 7** *Post-hoc* comparisons - effects on precision in participant 3

|  | D Means | t | P |
|---|---|---|---|
| **Session 1** |  |  |  |
| 2D vs. Direct | 66.5 | 0.96 | 0.334 NS |
| 2D vs. Fisheye | 227.7 | 3.57 | 0.000 |
| Fisheye vs. Direct | 354.8 | 5.11 | 0.000 |
| **Session 2** |  |  |  |
| 2D vs. Direct | 209.5 | 3.49 | 0.000 |
| 2D vs. Fisheye | 412.2 | 6.02 | 0.000 |
| Fisheye vs. Direct | 109.2 | 2.01 | 0.030 |
| **Session 3** |  |  |  |
| 2D vs. Direct | 269.4 | 2.53 | 0.020 |
| 2D vs. Fisheye | 267.7 | 2.21 | 0.020 |
| Fisheye vs. Direct | 1.6 | 0.01 | 0.985 NS |
| **Session 4** |  |  |  |
| 2D vs. Direct | 469.3 | 6.76 | 0.000 |
| 2D vs. Fisheye | 466.4 | 5.48 | 0.000 |
| Fisheye vs. Direct | 241.9 | 3.49 | 0.000 |
| **Session 5** |  |  |  |
| 2D vs. Direct | 522.1 | 7.52 | 0.000 |
| 2D vs. Fisheye | 70.2 | 1.09 | 0.300 NS |
| Fisheye vs. Direct | 420.9 | 6.48 | 0.000 |
| **Session 6** |  |  |  |
| 2D vs. Direct | 861.8 | 10.42 | 0.000 |
| 2D vs. Fisheye | 319.1 | 2.99 | 0.010 |
| Fisheye vs. Direct | 257.2 | 2.85 | 0.015 |
| **Session 7** |  |  |  |
| 2D vs. Direct | 387.3 | 6.57 | 0.000 |
| 2D vs. Fisheye | 393.9 | 6.23 | 0.000 |
| Fisheye vs. Direct | 60.6 | 0.07 | 0.938 NS |
| **Session 8** |  |  |  |
| 2D vs. Direct | 644.7 | 9.29 | 0.000 |
| 2D vs. Fisheye | 90.7 | 6.51 | 0.284 NS |
| Fisheye vs. Direct | 553.7 | 1.07 | 0.000 |

Results of the *post-hoc* comparisons for effects on precision of the three levels of 'viewing' ($V_3$) in the eight levels of 'session' ($S_8$) in level 3 of the 'participant' factor. Effect sizes (D Means), t values, and unadjusted probabilities (P) are given for each comparison the disadvantage of image-guidance may diminish with training and eventually level off, none of the individuals gets to perform as well as in the direct viewing condition in the last training sessions. In fact, the effects of the viewing conditions vary significantly between individuals as a function of the training session, as shown by the two-by-two interactions between these factors. The results of the relevant *post-hoc* comparisons, summarized in Tables 1, 2, 3, 4, 5, 6, 7 and 8, give a quantitative overview of these variations, which are difficult to interpret in terms of any simple explanation or model. Low-level explanations in terms of vision-proprioception conflict during task execution in the indirect viewing conditions would be a possible candidate. It has been shown that visual-proprioceptive matching, which is optimal in "natural" direct action viewing, is important for feeling in control of one's actions during the visual observation of one's own hand movements in eye-hand coordination tasks. This feeling of control, sometimes also referred to as *agency*, influences

**Table 8** *Post-hoc* comparisons - effects on time in participant 4

|  | D Means | t | P |
|---|---|---|---|
| Session 1 |  |  |  |
|   2D vs. Direct | 2.425 | 10.05 | 0.000 |
|   2D vs. Fisheye | 0.957 | 3.96 | 0.000 |
|   Fisheye vs. Direct | 3.381 | 14.02 | 0.000 |
| Session 2 |  |  |  |
|   2D vs. Direct | 2.202 | 9.13 | 0.000 |
|   2D vs. Fisheye | 0.217 | 0.90 | 0.368 NS |
|   Fisheye vs. Direct | 2.420 | 10.03 | 0.000 |
| Session 3 |  |  |  |
|   2D vs. Direct | 1.673 | 6.93 | 0.000 |
|   2D vs. Fisheye | 0.359 | 1.48 | 0.137 NS |
|   Fisheye vs. Direct | 2.031 | 8.42 | 0.000 |
| Session 4 |  |  |  |
|   2D vs. Direct | 1.772 | 7.34 | 0.000 |
|   2D vs. Fisheye | 0.177 | 0.73 | 0.464 NS |
|   Fisheye vs. Direct | 1.595 | 6.61 | 0.000 |
| Session 5 |  |  |  |
|   2D vs. Direct | 1.263 | 5.23 | 0.000 |
|   2D vs. Fisheye | 0.344 | 1.43 | 0.154 NS |
|   Fisheye vs. Direct | 1.607 | 6.66 | 0.000 |
| Session 6 |  |  |  |
|   2D vs. Direct | 1.686 | 6.99 | 0.000 |
|   2D vs. Fisheye | 0.243 | 1.01 | 0.314 NS |
|   Fisheye vs. Direct | 1.929 | 7.99 | 0.000 |
| Session 7 |  |  |  |
|   2D vs. Direct | 1.839 | 7.62 | 0.000 |
|   2D vs. Fisheye | 0.381 | 1.58 | 0.114 NS |
|   Fisheye vs. Direct | 2.220 | 9.20 | 0.000 |
| Session 8 |  |  |  |
|   2D vs. Direct | 1.740 | 7.21 | 0.000 |
|   2D vs. Fisheye | 0.068 | 0.28 | 0.778 NS |
|   Fisheye vs. Direct | 1.808 | 7.49 | 0.000 |

Results of the *post-hoc* comparisons for effects on time of the three levels of *'viewing'* ($V_3$) in the eight levels of *'session'* ($S_8$) in level 4 of the *'participant'* factor. Effect sizes (*D* Means), *t* values, and unadjusted probabilities (*P*) are given for each comparison

**Table 9** *Post-hoc* comparisons - effects on precision in participant 4

|  | D Means | t | P |
|---|---|---|---|
| Session 1 |  |  |  |
|   2D vs. Direct | 387.8 | 5.58 | 0.000 |
|   2D vs. Fisheye | 223.2 | 3.21 | 0.001 |
|   Fisheye vs. Direct | 164.5 | 2.37 | 0.018 |
| Session 2 |  |  |  |
|   2D vs. Direct | 365.6 | 8.49 | 0.000 |
|   2D vs. Fisheye | 105.8 | 1.53 | 0.126 NS |
|   Fisheye vs. Direct | 334.2 | 0.52 | 0.000 |
| Session 3 |  |  |  |
|   2D vs. Direct | 653.2 | 9.45 | 0.000 |
|   2D vs. Fisheye | 205.3 | 2.97 | 0.003 |
|   Fisheye vs. Direct | 448.4 | 6.48 | 0.000 |
| Session 4 |  |  |  |
|   2D vs. Direct | 393.8 | 5.69 | 0.000 |
|   2D vs. Fisheye | 12.57 | 0.18 | 0.856 NS |
|   Fisheye vs. Direct | 406.3 | 5.87 | 0.000 |
| Session 5 |  |  |  |
|   2D vs. Direct | 460.3 | 6.56 | 0.000 |
|   2D vs. Fisheye | 48.5 | 0.69 | 0.486 NS |
|   Fisheye vs. Direct | 412.1 | 5.96 | 0.000 |
| Session 6 |  |  |  |
|   2D vs. Direct | 539.9 | 6.353 | 0.000 |
|   2D vs. Fisheye | 100.4 | 5.627 | 0.146 NS |
|   Fisheye vs. Direct | 355.8 | 5.14 | 0.000 |
| Session 7 |  |  |  |
|   2D vs. Direct | 458.9 | 6.64 | 0.000 |
|   2D vs. Fisheye | 214.8 | 3.11 | 0.002 |
|   Fisheye vs. Direct | 244.4 | 3.53 | 0.000 |
| Session 8 |  |  |  |
|   2D vs. Direct | 281.1 | 4.06 | 0.000 |
|   2D vs. Fisheye | 45.9 | 0.66 | 0.507 NS |
|   Fisheye vs. Direct | 326.9 | 4.73 | 0.000 |

Results of the *post-hoc* comparisons for effects on precision of the three levels of *'viewing'* ($V_3$) in the eight levels of *'session'* ($S_8$) in level 4 of the *'participant'* factor. Effect sizes (*D* Means), *t* values, and unadjusted probabilities (*P*) are given for each comparison both the timing and the accuracy of hand movements [1]. Moreover, badly matched visual and proprioceptive inputs may reduce tactile sensitivity significantly [14]. We do, however, not think that this explanation is a likely candidate here. Firstly, although, compared with direct viewing, image viewing was not perfectly aligned with the forearm motor axis, it did not exceed the recommended maximal offset angle of 45°, beyond which performance may not be optimal (e.g. [35]). Moreover, previous work has shown that the direction of arm movements (vertical *vs* horizontal), not monitor position, matters critically in image-guided performance. Tasks requiring arm movements mostly in the vertical direction (as in our experimental task here) were performed faster and with more precision than tasks requiring essentially movements in the horizontal direction, regardless of where the monitor for viewing the video images was placed [10]. Secondly, the video images received from the camera in our experiment were professionally calibrated for both time and space. Spatial matching of the image conditions with the direct viewing condition was controlled by making sure the size of real-world action field parameters such as target, object, and tool sizes, were identical when viewed from the participants sitting position. Temporal matching was controlled by the algorithm driving the internal clock of the CPU, ensuring that the video-images where synchronized with the real-world actions, as specified in *Materials and Methods*. There was no perceptible mismatch or misalignment in either time or space between actions represented in the video-images and actions viewed directly. In motor learning, both low-level and high-level processes contribute to the evolution of performance with training (e.g. [42, 46]). High-level action intentions, which are closely linked to psychological factors such as response strategy preferences, were deliberately not controlled or selectively manipulated (no performance feed-back of any sort was given) in our experiment. "Natural" variations in high-level action intentions are therefore the most likely source of the inter-individual differences in the performances observed here. These typically occur spontaneously during training, are independent of low-levels task constraints, and reflect individual goal setting strategies predicted decades ago by results from seminal work in the field (e.g. [12]) and consistent with current neurophysiological models involving top-down decision control by the frontal lobe (e.g. [44]).

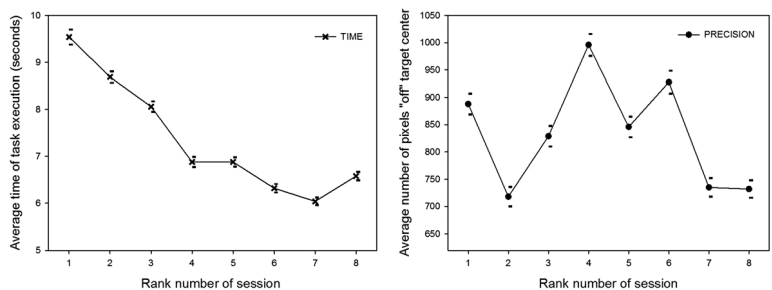

**Fig. 6** Average data for 'time' (*left*) and 'precision' (*right*) and their standard errors (SEMs), plotted as a function of the rank number of the experimental session. The effect of the 'session' factor is significant for both performance variables (see *'Analysis of variance'* in the Results section)

Wearing a glove does not significantly affect speed of execution, but does affect precision. This observation was not expected in the light of previous data (see [6]), but is explained by a reduction of tactile sensitivity to physical objects when no direct finger contact with the object is possible, which may be detrimental to feed-back signaling from hand to cortex for eye-hand coordination. This interpretation relates to earlier findings showing that the direct manipulation of objects by hand is combined with the visual and tactile integration of physical object parameters for action planning, gestural programming, and motor control ([18, 19, 8, 22]). This possibly involves cortical neurons with non-classic receptive field structures in the brain [56, 50, 52]. It can be assumed that under conditions of touch with direct contact between the physical object and the fingers of the hands, the finely tuned mechanoreceptors under the skin which control both fingertip forces and grasp kinematics [27] send stronger feed-back signals to these cortical neurons [31].

Tool-mediated object positioning was as precise as by-hand direct object positioning, but task execution was slower, as expected in the light of previous observations on novices (e.g.[55]). Tool-specific motor requirements (e.g. [11, 13, 19, 25, 32, 47]), such as having to grab and hold the handle of the tool, or having to adjust one's hand movements to the shape and the size of the tool, readily account for this effect. The effect of tool use on execution times is present throughout all the training sessions as shown in the conditional performance curves of the four individuals here.

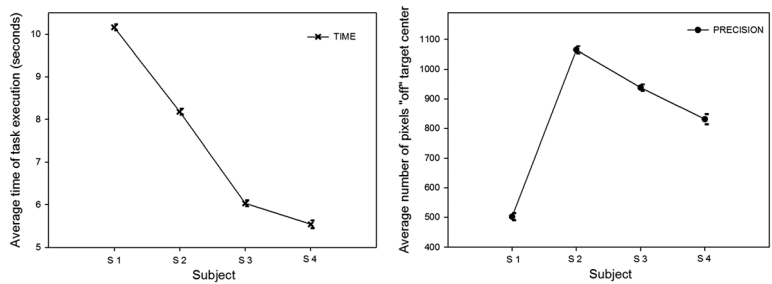

**Fig. 7** Average data for 'time' (*left*) and 'precision' (*right*) and their standard errors (SEMs), plotted for the four different participants. The effect of the 'participant' factor is significant for both performance variables and significantly interacts with the 'session' factor (see *'Analysis of variance'* in the Results section)

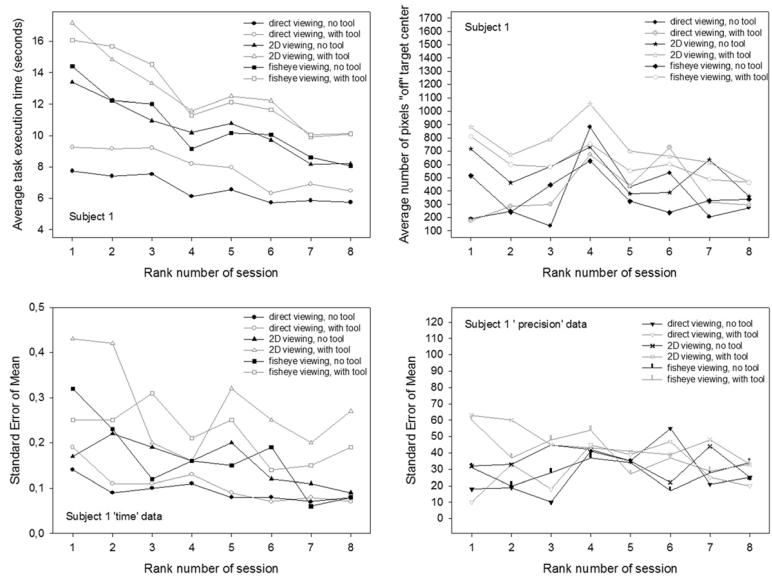

**Fig. 8** Conditional performance curves for 'time' and 'precision' for one participant (subject 1, female). Means (*upper panel*) and standard errors (*lower panel*) are plotted as a function of the rank number of the experimental training session

The most important results in the light of our study goal are the significant inter-individual differences in performance strategies during training found here in this image-guided pick-and-place task. These differences are reflected by strategy specific trade-offs between speed of task execution and the precision with which the object is placed on the targets. As predicted, these trade-offs occur spontaneously and without performance feed-back (e.g. [12]). The observations lead to understand why monitoring only execution times for learning curve ana-lysis in simulator training is not a viable option. Some trainees may get faster, but not necessarily better in the task, as shown here. Yet, in a majority of simulator train-ing programs for laparoscopic surgery, the relative

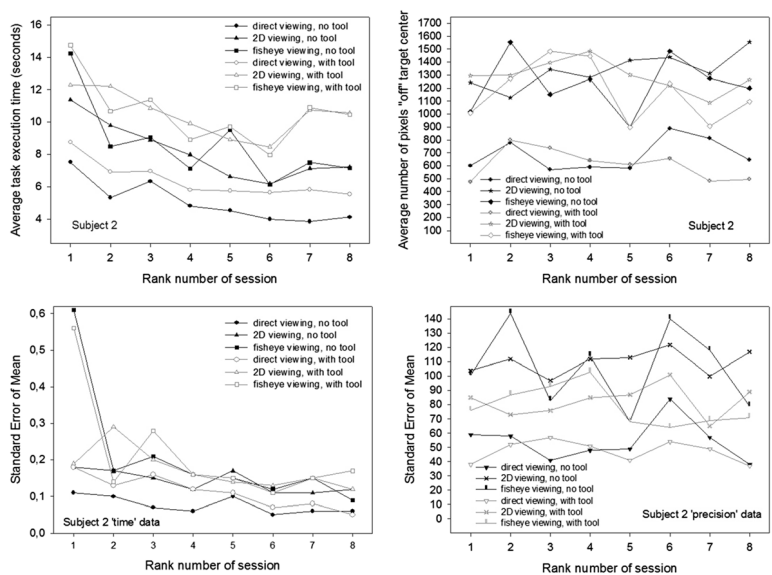

**Fig. 9** Conditional performance curves for 'time' and 'precision' for the second participant (subject 2, female). Means (*upper panel*) and standard errors (*lower panel*) are plotted as a function of the rank number of the experimental training session

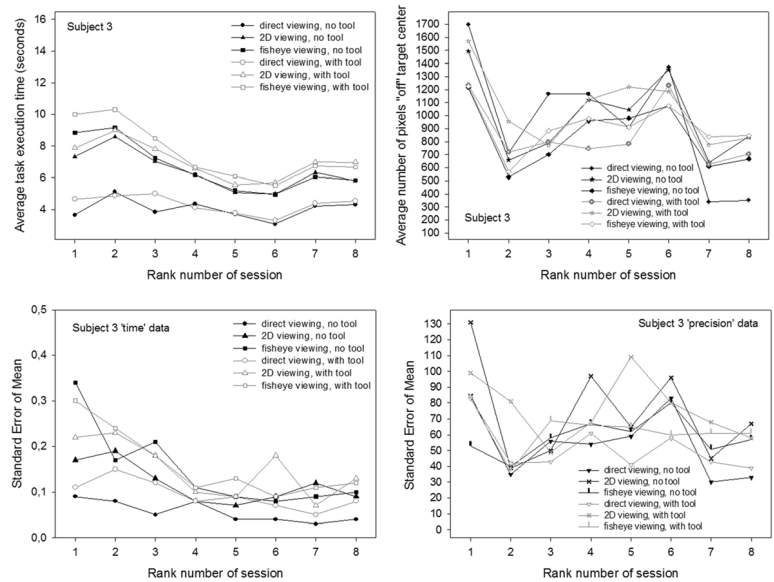

**Fig. 10** Conditional performance curves for 'time' and 'precision' of the third participant (subject 3, male). Means (*upper panel*) and standard errors (*lower panel*) are plotted as a function of the rank number of the experimental training session

precision of image-guided hand manoeuvres based on a conditional pixel-by-pixel analysis of hand or tool-movements from the video image data is not taken into account in the individual's learning curve. Neglecting the functional relationship between the time and the precision of task execution highlighted by the results from this study here is likely to have a cost. Individuals start the training sessions with different goals on their minds. Some place their effort on performing the object positioning task as fast as possible while others place their effort on being as precise as possible. The conditional performance curves reveal that the choice to privilege one strategy goal (either speed or precision) at the beginning has measurable consequences on the individual performance evolution at further stages of training. One trainee, who privileges precision at the outset (subject 1), becomes even more precise with further training, and also gets faster. Two other trainees (subjects 2 and 3) start fast, and re-adjust their execution times in mid-training, possibly because they realize that they may not perform with enough precision. One of them (subject 3) manages, indeed, to become more precise by adjusting his speed strategy to a slightly slower temporal performance level. One trainee, the fastest performer here (subject 4), starts fast and gets faster steadily with training in all conditions, yet, his precision never stabilizes. Even with twelve additional sessions, there was no measurable improvement in the precision score of this trainee.

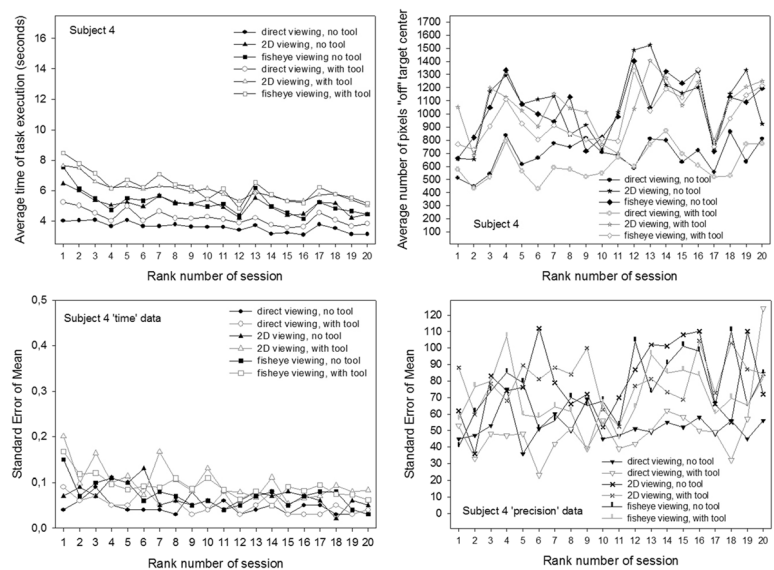

**Fig. 11** Conditional performance curves for 'time' and 'precision' of the fourth participant (subject 4, male). Means (*upper panel*) and standard errors (*lower panel*) are plotted as a function of the rank number of the experimental training session. This participant was run in twelve additional training sessions, producing a total of 20 sessions instead of eight

Experimental studies in the last century have proposed procedures for controlling a trainee's speed-accuracy trade-off in tasks where both time and precision matter critically. These procedures either aim at selectively rewarding either speed or precision during learning (e.g. [33]; for a more recent review see [44]). This can be achieved by providing adequate feed-back to the trainee, especially in the first training sessions. Making sure that the trainee gets as precise as possible before getting faster should be a priority in surgical simulator training. This can be achieved by instructing him/her to privilege accuracy rather than speed. Execution times then become faster automatically with training. Once a desired level of precision is reached by a trainee, time deadlines for task execution can be introduced, and progressively reduced during further training, to ensure the trainee will get as fast as possible without losing precision (e. g. [4, 5]). A major goal identified in recent analyses [17] is to ensure that the experimental evaluation of skills in surgical simulator scenarios is not subject to the development of a single observer bias over time, as may easily be the case in fully automated (unsupervised) skill rating procedures. Yet, these represent economy in manpower and are therefore likely to become the adopted standard, which will result in trainees not being coached individually and receiving no proper guidance on how to optimize their learning strategies. Supervised learning in small groups, in training loops with regular and adaptive skill assessment, as shown here in Figure 12, represents a better and not necessarily more costly alternative in the light of the findings reported here, especially in surgical simulator training, where reliable performance standards are urgently needed.

## Conclusions

The results from this study reveal complex and spontaneously occurring trade-offs between time and precision in the performance of four individuals, all absolute beginners, in visual spatial learning of an image-guided object positioning task. These trade-offs reflect cognitive strategy variations that need to be monitored individually to ensure effective skill learning. Collecting only time data to establish learning curves is not an option, as getting faster does not straightforwardly imply getting better at the task. Training procedures should include skill evaluation by expert psychologists and procedures for the adaptive control of speed-accuracy trade-offs in the performances of novices.

### Acknowledgements
Non applicable.

### Funding
The study was funded by the Initiative D'EXcellence (IDEX) of the University of Strasbourg. Material for building the experimental platform was financed by CNRS (Appels à Projets Interdisciplinaires 2015 to BDL). The funding bodies had no role in the study design, data collection, or decision to submit this manuscript for publication.

### Availability of data and materials
All data are fully displayed, as graphs or tables, in the Results section of this manuscript. The authors share the raw via a publicly available repository at: https://osf.io/rsh9/

### Authors' information
Laboratoire ICube, UMR 7357 CNRS-Université de Strasbourg, FRANCE

### Competing interests
The authors declare that they have no competing interests of any nature, financial or non-financial.

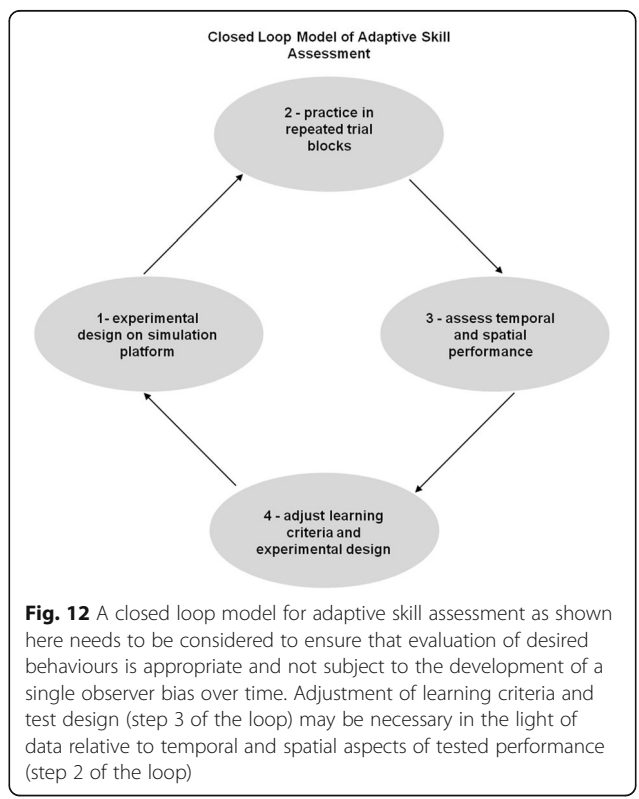

**Fig. 12** A closed loop model for adaptive skill assessment as shown here needs to be considered to ensure that evaluation of desired behaviours is appropriate and not subject to the development of a single observer bias over time. Adjustment of learning criteria and test design (step 3 of the loop) may be necessary in the light of data relative to temporal and spatial aspects of tested performance (step 2 of the loop)

**Consent for publication**
Consent to publish the images in Fig. 1 has been obtained from the individual shown in these images.

**Ethics approval and consent to participate**
The study was conducted in conformity with the Helsinki Declaration relative to scientific experiments on human individuals with the full approval of the ethics board of the corresponding author's host institution (CNRS). All participants were volunteers and provided written informed consent to participate.

**References**
1. Balslev D, Cole J, Miall RC. Proprioception contributes to the sense of agency during visual observation of hand movements: evidence from temporal judgments of action. J Cogn Neurosci. 2007;19:1535–41.
2. Batmaz AU, de Mathelin M, Dresp-Langley B. Effects of indirect screen vision and tool-use on the time and precision of object positioning on real-world targets. Perception. 2016;45(ECVP Supplement):196.
3. Bogner MS. Human error in medicine. New Jersey: Lawrence Erlbaum Associates, Hillsdale; 1994.
4. Bonnet C. Psychophysical approaches, contextual effects, and response bias. In: Caverni JP, Fabre JM, Gonzales M, editors. Cognitive Biases. Amsterdam: Elsevier; 1990. p. 309–18.
5. Bonnet C, Dresp B. A fast procedure for studying conditional accuracy functions. Behav Res Instrum Comput. 1993;25:2–8.
6. Desai S, Konz S. Tactile inspection performance with and without gloves. Proceedings of the human factors and ergonomics society annual meeting. 1983;27:782–85.
7. Det MJ, Mijerink WJHJ, Hoff C, Totté ER, Pierie JPEN. Optimal ergonomics for laparoscopic surgery in minimally invasive surgery suites: a review and guidelines. Surg Endosc. 2009;23:1279–85.
8. Di Pellegrino G, Ladavas E, Farnè A. Seeing where your hands are. Nature. 1997;388:730.
9. Dresp-Langley B. Principles of perceptual grouping: implications for image guided surgery. Front. Psychol. 2015;6:1565.
10. Emam TA, Hanna G, Cuschieri A. Ergonomic principles of task alignment, visual display, and direction of execution of laparoscopic bowel suturing. Surg Endosc. 2002;16:267–71.
11. Farnè A, Ladavas E. Dynamic size-change of hand peripersonal space following tool use. Neuroreport. 2000;11:1645–9.
12. Fitts PM. The information capacity of the human motor system in controlling the amplitude of movement. J Exp Psychol. 1954;47:381–91.
13. Fogassi L, Gallese V. Action as a binding key to multisensory integration. In: Calvert G, Spence C, Stein BE, editors. Handbook of multisensory processes. Cambridge, MA: MIT Press; 2004. p. 425–41.
14. Folegatti A, de Vignemont F, Pavani F, Rossetti Y, Farnè A. Losing one's hand: visual-proprioceptive conflict affects touch perception. PLos One. 2009;4(9):e6920.
15. Gallace A, Spence C. The cognitive and neural correlates of "tactile consciousness": A multisensory perspective. Conscious Cogn. 2008;17:370–407.
16. Gallagher AG, Ritter EM, Lederman AB, McClusky 3rd DA, Smith CD. Video-assisted surgery represents more than a loss of three-dimensional vision. Am J Surg. 2005;189:76–80.
17. Gallagher A. G, O'Sullivan G. C. Fundamentals in surgical simulation: principles and practice. Improving medical outcome - zero tolerance series. Apell P. editor. Springer Business Media, e-book;2011. http://www.springer.com/us/book/9780857297624.
18. Gibson JJ. Observations on active touch. Psychol. Rev. 1962;69:477–91.
19. Graziano MS, Cross CG. The representation of extrapersonal space: a possible role for bimodal, visual-tactile neurons. In: Gazzaniga MS, editor. The Cognitive Neurosciences. Cambridge, MA: MIT Press; 1994. p. 1021–34.
20. Hanna GB, Shimi SM, Cuschieri A. Task performance in endoscpic surgery is influenced by location of the image display. Ann Surg. 1998;4:481–4.
21. Haveran LA, Novitsky YW, Czerniach DR, Kaban GK, Taylor M, Gallagher-Dorval K, Schmidt R, Kelly JJ, Litwin DEM. Optimizing laparoscopic task efficiency: the role of camera and monitor positions. Surg Endosc. 2007;21:980–4.
22. Held R. Visual-haptic mapping and the origin of crossmodal identity. Optom Vis Sci. 2009;86:595–8.
23. Henriques DY, Cressman EK. Visuo-motor adaptation and proprioceptive recalibration. J Mot Behav. 2012;44:435–44.
24. Huang VS, Mazzoni PP, Krakauer JW. Rethinking motor learning and savings in adaptation paradigms: model-free memory for successful actions combines with internal models. Neuron. 2011;70:787–801.
25. Humphreys GW, Riddoch MJ, Forti S, Ackroyd K. Action influences spatial perception: Neuropsychological evidence. Vis Cogn. 2004;11:401–27.
26. Jalote-Parmar A, Badke-Schaub P, Ali W, Samset E. Cognitive processes as integrative component for developing expert decision-making systems: a workflow centered framework. J Biomed Inform. 2010;43:60–74.
27. Jenmalm P, Dahlstedt S, Johansson RS. Visual and tactile information about object curvature control fingertip forces and grasp kinematics in human dexterous manipulation. J Neurophysiol. 2000;84:2984–97.
28. Kanhere A, Aldalali B, Greenberg JA, Heise CP, Zhang L, Jiang H. Reconfigurable micro-camera array with panoramic vision for surgical imaging. J Micro Electromechanical Syst. 2013;22:1057–7157.
29. Krakauer JW, Mazzoni P. Human sensorimotor learning: adaptation, skill, and beyond. Curr Opin Neurobiol. 2011;21:636–44.
30. Kumler J. J, Bauer M. L. Fisheye lens designs and their relative performance, in Proc. SPIE, San Diego, CA. 2000; 4093:360–9.
31. Lamotte RH, Friedman RM, Lu C, Khalsa PS, Srinivasan MA. Raised object on a planar surface stroked across the fingerpad: responses of cutaneous mechanoreceptors to shape and orientation. J Neurophysiol. 1998;80:2446–66.
32. Longo MR, Lourenco SF. On the nature of near space: Effects of tool use and the transition to far space. Neuropsychologia. 2006;44:977–81.
33. Luce RD. Response times: Their role in inferring elementary mental organization. New York: Oxford University Press; 1986.
34. Luger T, Bosch T, Hoozemans M, de Looze M, Veeger D. Task variation during simulated, repetitive, low-intensity work: influence on manifestations of shoulder muscle fatigue, perceived discomfort and upper-body postures. Ergonomics. 2015;58:1851–67.
35. Maithel SK, Villegas L, Stylopoulos N, Dawson S, Jones DB. Simulated laparoscopy using a head-mounted display vs traditional video monitor: an assessment of performance and muscle fatigue. Surg Endosc. 2005;19:406–11.
36. Maravita A, et al. Reaching with a tool extends visual-tactile interactions into far space: evidence from crossmodal extinction. Neuropsychologia. 2001;39:580–5.
37. Maravita A, Ikiri A. Tools for the body (schema). Trends Cogn Sci. 2004;8:79–86.
38. McClelland JL. On the time relations of mental processes: an examination of systems of processes in cascade. Psychol Rev. 1979;86:375–407.
39. Meyer DE, Irwin A, Osman AM, Kounios J. The dynamics of cognition and action: mental processes inferred from speed-accuracy decomposition. Psychol Rev. 1988;95:183–237.
40. Oldfield RC. The assessment and analysis of handedness: the Edinburgh inventory. Neuropsychologia. 1971;9:97–113.
41. Ollmann RT. Choice reaction time and the problem of distinguishing task effects from strategy effects. In: Domic S, editor. Attention & performance VI. Hillsdale: Erlbaum; 1977. p. 99–113.
42. Preston C, Newport R. Self-denial and the role of intentions in the attribution of agency. Conscious Cogn. 2010;19:986–98.
43. Sarlegna F, Blouin J, Bresciani JP, Bourdin C, Vercher JL, Gauthier GM. Target and hand position information in the online control of goal-directed arm movements. Exp Brain Res. 2003;151:524–35.
44. Schall JD, Stuphorn V, Brown JW. Monitoring and control of action by the frontal lobes. Neuron. 2002;36:309–22.
45. Schout BMA, Hendrix AJM, Scheele F, Bemelmans BLH, Scherpbier AJJA. Validation and implementation of surgical simulators: a critical review of present, past, and future. Surg Endosc. 2010;24:536–46.
46. Slachevsky A, Pillon B, Fourneret P, Pradat-Diehl P, Jeannerod M, Dubois B. Preserved adjustments but impaired awareness in a sensori-motor conflict following prefrontal lesions. J Cogn Neurosci. 2001;13:332–40.
47. Sommer R. Personal space: the behavioral basis of design. Englewood Cliffs: Prentice-Hall; 1969.
48. Spence C, Nicholls MER, Driver J. The cost of expecting events in the wrong sensory modality. Percept Psychophys. 2001;63:330–6.
49. Solomonczyk D, Cressman EK, Henriques DY. The role of the cross-sensory error signal in visuo-motor adaptation. Exp Brain Res. 2013;228:313–25.

50. Stein BE, Wallace MW, Stanford TR, Jiang W. Cortex governs multisensory integration in the midbrain. Neuroscientist. 2002;8:306–314.
51. Stüdeli T, Freudenthal A, de Ridder H. Evaluation framework of ergonomic requirements for iterative design development of computer systems and their user interfaces for minimal invasive therapy. In: Toomingas A, Lantz A, Berns T, editors. Proceedings WWCS 2007 Computing Systems for Human Benefits, 8th International Conference on work with Computing Systems, May 21–24, Stockholm, Sweden; 2007.
52. Thakur PH, Fitzgerald PJ, Hsiao SS. Second-order receptive fields reveal multi-digit interactions in area 3B of the macaque monkey. J Neurophysiol. 2012;108:243–62.
53. Uhrich ML, Underwood RA, Standeven JW, Soper NJ, Engsberg JR. Assessment of fatigue, monitor placement, and surgical experience during simulated laparoscopic surgery. Surg Endosc. 2002;16:635–9.
54. Verdaasdonk J, Dankelman J, Lange JF, Stassen LP. Incorporation of proficiency criteria for basic laparoscopic skills training: how does it work? Surg Endosc. 2008;22:2609–15.
55. Wilson MR, Mc Grath JS, Vine SJ, Brewer J, Defriend D, Masters RSW. Perceptual impairment and psychomotor control in virtual laparoscopic surgery. Surg Endosc. 2011;25:2268–74.
56. Zangaladze A, Epstein CM, Grafton ST, Sathian K. Involvement of visual cortex in tactile discrimination of orientation. Nature. 1999;401:587–590.